\documentclass[aps,prb,amsmath,twocolumn,showpacs,superscriptaddress,showkeys,10pt]{revtex4-1}
\usepackage{graphicx} 

\newcommand{\eff}{\text{eff}}
\newcommand{\AFM}{\text{AFM}}

\begin{document}

\title{Quasi-two-dimensional $S=1/2$ magnetism of Cu[C$_6$H$_2$(COO)$_4$][C$_2$H$_5$NH$_3$]$_2$}

\author{R. Nath}
\email{rnath@iisertvm.ac.in}
\affiliation{School of Physics, Indian Institute of Science
Education and Research Thiruvananthapuram-695016, Kerala, India}

\author{M. Padmanabhan}
\affiliation{School of Chemistry, Indian Institute of Science 
Education and Research Thiruvananthapuram-695016, Kerala, India}
\affiliation{School of Chemical Sciences, Mahatma Gandhi University Kottayam-686560, Kerala, India}

\author{S. Baby}
\affiliation{School of Chemical Sciences, Mahatma Gandhi University Kottayam-686560, Kerala, India}

\author{A. Thirumurugan}
\affiliation{School of Chemistry, Indian Institute of Science 
Education and Research Thiruvananthapuram-695016, Kerala, India}

\author{D. Ehlers}
\author{M. Hemmida}
\author{H.-A. Krug~von~Nidda}
\affiliation{Experimentalphysik V, Center for Electronic Correlations and Magnetism, Institute for Physics, Augsburg University, 86135 Augsburg, Germany}

\author{A. A. Tsirlin}
\email{altsirlin@gmail.com}
\affiliation{National Institute of Chemical Physics and Biophysics, 12618 Tallinn, Estonia}

\date{\today}

\begin{abstract}
We report structural and magnetic properties of the \mbox{spin-$\frac12$} quantum antiferromagnet Cu[C$_6$H$_2$(COO)$_4$][C$_2$H$_5$NH$_3$]$_2$ by means of single-crystal x-ray diffraction, magnetization, heat capacity, and electron spin resonance (ESR) measurements on polycrystalline samples, as well as band-structure calculations. The triclinic crystal structure of this compound features CuO$_4$ plaquette units connected into a two-dimensional framework through anions of the pyromellitic acid [C$_6$H$_2$(COO)$_4$]$^{4-}$. The ethylamine cations [C$_2$H$_5$NH$_3]^+$ are located between the layers and act as spacers. Magnetic susceptibility and heat capacity measurements establish a quasi-two-dimensional, weakly anisotropic and non-frustrated spin-$\frac12$ square lattice with the ratio of the couplings $J_a/J_c\simeq 0.7$ along the $a$ and $c$ directions, respectively. No clear signatures of the long-range magnetic order are seen in thermodynamic measurements down to 1.8\,K. However, the gradual broadening of the ESR line suggests that magnetic ordering occurs at lower temperatures. Leading magnetic couplings are mediated by the organic anion of the pyromellitic acid and exhibit a non-trivial dependence on the Cu--Cu distance, with the stronger coupling between those Cu atoms that are further apart.
\end{abstract}
\maketitle

\section{Introduction}
Cu$^{2+}$ compounds with organic cations and anions are in the focus of current research on quantum magnetism. Their advantages include facile crystal growth from the solution\cite{yankova2012} and large Cu--Cu separations leading to relatively weak exchange couplings that are on the scale of feasible magnetic fields and reveal remarkable sensitivity to the applied pressure. Therefore, both magnetic field and external pressure can be used to change the physical regime of the system and tune it toward a new phase or a quantum critical point. Remarkable examples include the operational low-temperature magnetocaloric effect in the Cu-oxalate-based compound,\cite{wolf2011} magnetic-field-induced ferroelectricity in sulfolane copper chloride Sul-Cu$_2$Cl$_4$,\cite{schrettle2013} and pressure-induced incommensurate magnetism in piperazinium copper chloride PHCC.\cite{thede2014}

Despite numerous experimental studies, microscopic aspects of Cu$^{2+}$ magnets with organic components are relatively less developed.\cite{[{Note, however, several recent studies: }][{}]jornet-somoza2010a,*jornet-somoza2010b,vela2013} Many of these systems are easy to understand empirically, because organic molecules provide only a few linkages between the spin-$\frac12$ Cu$^{2+}$ ions, hence forming a clearly identifiable backbone of the low-dimensional magnetic unit.\cite{goddard2008,manson2011,goddard2012} Nevertheless, detailed understanding of the underlying exchange mechanisms is vitally important for the deliberate preparation of new compounds. Moreover, as we show below, the trends in magnetic exchange through organic molecules are far from being trivial and extend our knowledge of superexchange interactions in general.

Here, we consider Cu(PM)(EA)$_2$, where EA stands for the ethylamine [C$_2$H$_5$NH$_3]^+$ cation, and PM is the [C$_6$H$_2$(COO)$_4$]$^{4-}$ anion of pyromellitic acid. This newly synthesized compound features layered crystal structure, with organic anions connecting Cu$^{2+}$ ions into a two-dimensional (2D) square-lattice-like network. Two EA$^+$ cations then balance the negative charge of the resulting \textit{anionic} framework and reside between the layers. This type of structure is clearly reminiscent of quasi-2D magnets Cu(pz)$_2$X$_2$, where Cu$^{2+}$ ions are linked through pyrazine molecules (pz) and form a \textit{cationic} framework. Its charge is compensated by inorganic anions X, such as ClO$_4^-$, F$^-$, etc.\cite{goddard2008,lancaster2007} 

An interesting feature of Cu(pz)$_2$X$_2$ and related compounds is the weak frustration\cite{tsyrulin2009,*tsyrulin2010,siahatgar2011} of their square-lattice magnetic network by second-neighbor interactions $J_2$ yielding the well-known model of the frustrated square lattice (FSL) that enjoys close theoretical attention\cite{[{}][{, and references therein}]fsl2004,*tsirlin2009} and possible connections to high-temperature superconductivity in doped Cu$^{2+}$ oxides.\cite{lee2008} We thus expected that Cu(PM)(EA)$_2$ might also show the FSL physics and reveal a stronger frustration than in Cu(pz)$_2$X$_2$. The latter compounds feature two nearest-neighbor couplings mediated by different pyrazine molecules, so that a direct superexchange pathway for $J_2$ is missing. In contrast, both first- and second-neighbor couplings in Cu(PM)(EA)$_2$ should be mediated by the same organic molecule of the pyromellitic acid, hence an increase in $J_2$ is naturally expected.

Our experimental data and microscopic analysis confirm the quasi-2D nature of Cu(PM)(EA)$_2$ with a weak spatial anisotropy of in-plane magnetic couplings and a very small interlayer coupling. We do not find any signatures of the frustration, though. Nevertheless, our data disclose a non-trivial mechanism of the remarkably long-range superexchange between the Cu$^{2+}$ ions. This superexchange is mediated by the carbon atoms involved in the phenyl ring of the pyromellitic acid. Its implications for other quantum magnets are discussed.       

\begin{figure}
\includegraphics{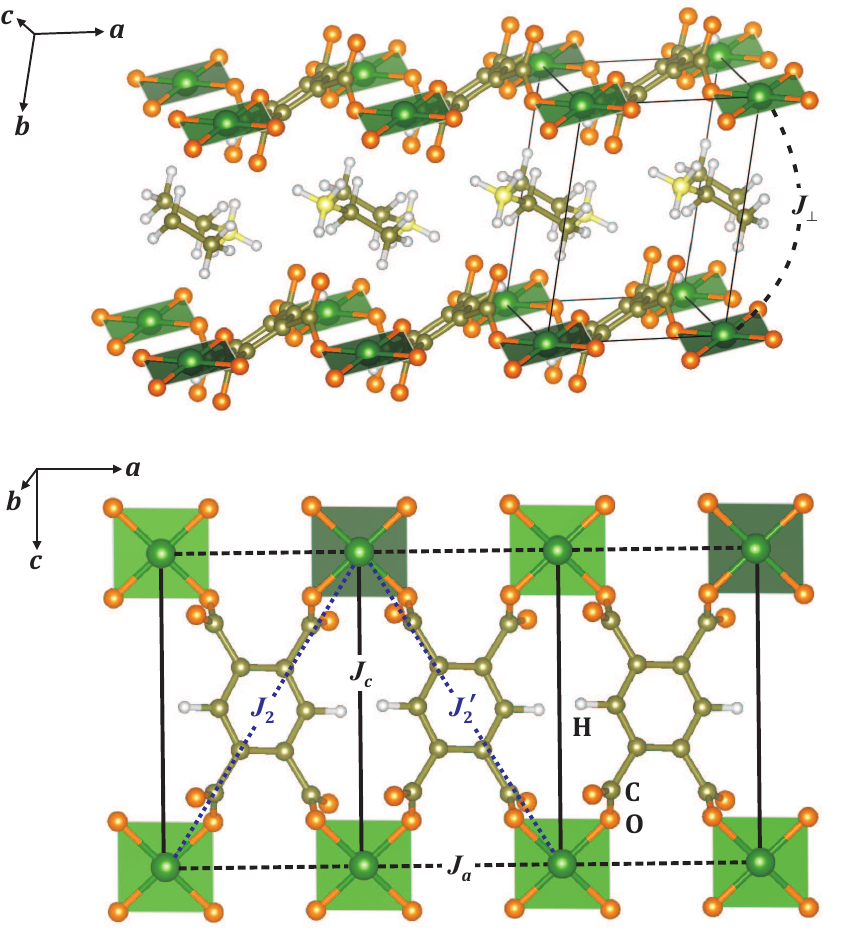}
\caption{\label{structure} Top panel: an overall view of the Cu(PM)(EA)$_2$ structure featuring negatively charged [Cu(PM)]$^{2-}$ layers interleaved by the EA$^+$ cations. Bottom panel: the structure of the [Cu(PM)]$^{2-}$ layers and relevant magnetic couplings forming a rectangular $J_a-J_c$ lattice of spin-$\frac12$ Cu$^{2+}$ ions with frustrating diagonal couplings $J_2$ and $J_2'$. The estimates of individual exchange couplings are given in Table~\ref{tab:exchanges}. The drawing was prepared using the \texttt{VESTA} software.\cite{vesta}
}
\end{figure}

\section{Methodology}
For the preparation of single crystals of Cu(PM)(EA)$_2$, an aqueous solution of Cu(CH$_3$COO)$_2\cdot$H$_2$O (5\,mM, 1.0\,g) was treated with 2 equivalents of ethylamine (0.9\,ml, 10\,mM, 70\% solution in water) followed by the addition of pyromellitic (1,2,4,5-benzenetetracarboxylic) acid (5\,mM, 1.27\,g) solution in dimethylformamide. The initially formed greenish-blue precipitate was filtered out. The ensuing clear light-blue solution was kept at room temperature for slow evaporation. Blue needle-shaped crystals of the title compound were obtained after 8 days. They were repeatedly washed with water and finally with methanol. The sample dried in air was found to be phase-pure form of Cu(PM)(EA)$_2$. Yield: 50\% (based on Cu). Analysis (calculated for C$_{14}$H$_{18}$CuN$_2$O$_8$):  C, 41.39; H, 4.43 ; N, 6.89. Found: C, 41.67; H, 4.42; N, 7.04\%. Infra-red data (KBr pellet, in~cm$^{-1}$): 3154 br, 3092 br, 3056 s, 2934 br, 2911 br, 2871 s, 2809 s, 1630 s, 1589 m,  1497 m, 1429 m, 1360 s, 1316 m, 1196 w, 1140 w, 1040 m, 992 s, 820 m, 712 s, 690 s, 536 s, 494 s.

Single crystal x-ray diffraction (Bruker APEX-II machine with MoK$_{\alpha1}$ radiation of wave length $\lambda = 0.71073$\,\r A) was performed on a high-quality single crystal of Cu(PM)(EA)$_2$ at room temperature. The data were reduced using SAINTPLUS,\cite{smart2004} and an empirical absorption correction was applied using the SADABS program.\cite{sheldrick1994} The crystal structure was solved by direct methods using SHELXS97 and refined using SHELXL97 from the WinGx suite of programs (Version 1.63.04a).\cite{sheldrick1997,*sheldrick1997b} All the hydrogen atoms were placed geometrically and held in the riding mode for the final refinements. The final refinements included atomic positions for all the atoms, anisotropic thermal parameters for all the non-hydrogen atoms and isotropic thermal parameters for the hydrogen atoms.
The crystal data and structure refinement parameters are shown in Table~\ref{crystaldata}. Few single crystals were crushed into powder, and powder x-ray diffraction (PANalytical machine with CuK$_{\alpha}$ radiation of wave length $\lambda = 1.54060$\,\r A) was performed to confirm the purity of polycrystalline samples. Unfortunately, the size of individual single crystals of Cu(PM)(EA)$_2$ was insufficient for thermodynamic measurements.

Magnetic susceptibility ($\chi$) was measured on the powder sample as a function of temperature (1.8~K~$\leq T \leq$~300~K) and at different applied magnetic fields ($H$) using a SQUID-VSM (Quantum Design). The magnetization isotherm ($M$ vs. $H$) was measured at $T = 2.5$\,K in static fields up to 14\,T with the VSM and in pulsed magnetic fields up to 30\,T at the Dresden High Magnetic Field Laboratory (HLD). Heat capacity ($C_p$) was measured with Quantum Design PPMS as a function of $T$ and $H$ on three crystalline needles glued together on the heat capacity platform.

Electron spin resonance (ESR) measurements were carried out in a Bruker ELEXSYS E500-CW spectrometer working at X-band (9.4 GHz) frequencies equipped with a continuous-flow $^{4}$He cryostat \emph{Oxford Instruments} ESR 900 and ESR 910 covering the temperature range $1.8~{\rm K} \leq T \leq 300$~K. Due to the lock-in amplification with field modulation, the ESR spectra record the
field derivative of the microwave absorption dependent on the external static field.
For this purpose the samples were fixed in \emph{Suprasil} quartz-glass tubes with paraffin.

Individual exchange couplings in Cu(PM)(EA)$_2$ were evaluated by density-functional (DFT) band-structure calculations in the \texttt{FPLO} code.\cite{fplo} The Perdew-Burke-Ernzerhof (GGA) flavor of the exchange-correlation potential\cite{pbe96} was supplied with the mean-field GGA+$U$ correction for strong electronic correlations in the Cu $3d$ shell using the on-site Coulomb repulsion $U_d=9.5$\,eV and Hund's exchange $J_d=1$\,eV, as applied in previous studies.\cite{janson2012,nath2013} All calculations were performed for the experimental crystal structure with the positions of hydrogen atoms fully relaxed within GGA.\cite{[{Note that the relaxation within GGA is sufficient for this purpose, as shown in: }][{}]lebernegg2013} Thermodynamic properties in zero field and in applied magnetic fields were calculated numerically using the \texttt{loop}\cite{loop} and \verb|dirloop_sse|\cite{dirloop} quantum Monte-Carlo (QMC) algorithms of the \texttt{ALPS} simulations package.\cite{alps}

\section{\textbf{Results}}
\subsection{Crystal Structure}
\label{sec:structure}
\begin{table}
\caption{\label{crystaldata} Crystal structure data for Cu(PM)(EA)$_2$ at room temperature.}
\begin{ruledtabular}
\begin{tabular}{ccccccc}
 Empirical formula & C$_{14}$H$_{18}$CuN$_2$O$_8$ \\
 Formula weight & 405.84 \\
 Temperature & 293~K \\
 Wave length ($\lambda$) & 0.71073~\AA \\
 Crystal system & Triclinic \\ 
 Space group & $P\bar 1$ \\
 Lattice parameters & $a=5.8610(1)$~\AA \\
                     & $b=8.3614(2)$~\AA \\
                     & $c=9.1772(2)$~\AA \\
                     & $\alpha=63.387(1)^{\circ}$ \\  
                     & $\beta=89.913(1)^{\circ}$ \\
                     & $\gamma=76.531(1)^{\circ}$ \\
 Volume ($V$) & 388.29(2)~\r A$^3$ \\
     $Z$      & 1 \\
 Calculated density ($\rho_{\rm cal}$) & 1.736~mg/mm$^3$ \\
 Absorption coefficient ($\mu$) & 1.455~mm$^{-1}$ \\
 F(000) &	209.0 \\
 Crystal size &	0.2 $\times$ 0.15 $\times$ 0.1 mm$^3$ \\
 2$\Theta$ range for data collection &	5.556$^{\circ}$ to 56.646$^{\circ}$ \\
 Index ranges &	$-7 \leq h \leq 7$, \\
              & $-11 \leq k \leq 11$, \\
              & $-12 \leq l \leq 12$ \\
 Reflections collected &	6870 \\
 Independent reflections &	1919 [$R_{\rm int} = 0.0186$] \\
 Data/restraints/parameters &	1919/0/117 \\
 Goodness-of-fit on F$^2$ &	1.278 \\
 Final $R$ indexes [$I\geq2\sigma(I)$] &	$R_1 = 0.0195$, $wR_2 = 0.0634$ \\
 Final $R$ indexes [all data] &	$R_1 = 0.0222$, $wR_2 = 0.0782$ \\
 Largest diff. peak/hole & 	$+0.60/-0.63$~$e$\,\r A$^{-3}$ \\
\end{tabular}
\end{ruledtabular}
\end{table}

Crystals of Cu(PM)(EA)$_2$ feature triclinic symmetry, space group $P\bar 1$. Their lattice parameters, atomic positions, and main interatomic distances and angles are given in Tables~\ref{crystaldata}, \ref{atom}, and~\ref{distance}, respectively. The Cu$^{2+}$ ions are at the inversion center in the origin of the unit cell. They form nearly flat CuO$_4$ plaquette units that are linked by the PM$^{4-}$ anions and build layers in the $ac$ plane. The EA$^+$ cations are located between these layers and connected to the anionic framework [Cu(PM)]$^{2-}$ through multiple hydrogen bonds. 

A simple visual examination of the crystal structure suggests a pronounced spatial anisotropy. Given the large distance and the lack of direct connections between the CuO$_4$ plaquettes along the crystallographic $b$-direction, magnetic couplings along this direction should be very weak. The PM$^{4-}$ anion linking the Cu$^{2+}$ ions may induce four different couplings in the $ac$ plane: the nearest-neighbor exchanges $J_a$ (along $[100]$) and $J_c$ ($[001]$) and second-neighbor diagonal exchanges $J_2$ ($[10\bar 1]$) and $J_2'$ ($[101]$), all running through the benzene ring of the PM molecule. The resulting model is a spatially anisotropic frustrated square lattice akin to those considered in Refs.~\onlinecite{tsirlin2009,schmidt2010,*schmidt2011,bishop2008,majumdar2010}. According to Cu-Cu distances in the experimental crystal structure (Table~\ref{tab:exchanges}), we expect $J_a>J_c>J_2\simeq J_2'$, but experimentally and microscopically, the order of couplings turns out to be different: $J_c>J_a\gg J_2$, $J_2'$. In the following, we study the spin lattice of Cu(PM)(EA)$_2$ and the origin of magnetic superexchange in this compound.

\begin{table}
\caption{\label{atom} Fractional atomic coordinates ($\times 10^{4}$) for Cu(PM)(EA)$_2$. The isotropic atomic displacement parameters (ADP) $U_{\rm eq}$ (in 10$^{-2}$\,\r A$^{-3}$) are defined as one-third of the trace of the orthogonal $U_{ij}$ tensor. The error bars are from the least-squares structure refinement. The positions of hydrogen atoms were additionally relaxed in a DFT calculation, hence the error bars are not given, and the relevant atom connected to hydrogen is listed instead of the ADP.}
\begin{ruledtabular}
\begin{tabular}{ccccccc}
 Atoms & $x/a$ & $y/b$ & $z/c$ & $U_{\rm eq}$ \\ \hline
 Cu1 & 0 & 0 & 0 & 11.46(10) \\
 O1 & 1999.0(19) & 786.7(16) & 1165.1(13) & 17.2(2) \\
 O2	& 2850(2)	& $-2243.7(17)$	& 2823.6(15) & 24.5(3) \\
 O3	& 6970(2)	& 2913.1(17)	& 1094.5(15)	& 25.0(3) \\
 O4	& 7557(2)	& 27.5(17)	& 1407.8(14) & 19.4(2) \\
 N1	& 6718(3)	& 6088(2)	& 1595.4(19)	& 26.1(3) \\
 C1	& 2948(2)	& $-654(2)$	& 2502.5(18)	& 14.5(3) \\
 C2	& 4096(2)	& $-335.2(19)$	& 3782.4(17) & 12.5(3) \\
 C3	& 5819(2)	& 650.4(19)	& 3456.3(17)	& 12.4(3) \\
 C4	& 6847(2)	& 1292(2)	& 1847.9(17)	& 13.9(3) \\
 C5	& 3301(2)	& $-981(2)$	& 5323.3(17)	& 14.2(3) \\
 C6	& 10885(4) & 4449(3) & 2683(3) & 39.3(5) \\
 C7	& 8579(3)	& 5448(2)	& 2985(2)	& 27.7(4) \\
 H1 & 7123    & 7091    & 521     & N1 \\
 H2 & 5099    & 6686    & 1851    & N1 \\
 H3 & 6556    & 4975    & 1396    & N1 \\
 H4 & 1928    & 8278    & 5562    & C1 \\
 H5 & 7922    & 4559    & 4108    & C4 \\
 H6 & 8742    & 6683    & 3075    & C4 \\
 H7 & 1519    & 5378    & 1568    & C5 \\
 H8 & 704     & 3249    & 2537    & C5 \\
 H9 & 2238    & 3960    & 3720    & C5 \\
\end{tabular}
\end{ruledtabular}
\end{table}

\begin{table}
\caption{\label{distance} Selected bond distances and angles for Cu(PM)(EA)$_2$.\footnote{Symmetry indices are defined as follows:\\ $^1$ $-x,\,-y,\,-z$\\ $^2$ $-1\!-\!x,\,y,\,z$\\ $^3$ $1\!-\!x,\,-y,\,-z$\\ $^4$ $1\!-\!x,\,-y,\,1\!-\!z$\\ $^5$ $1\!+\!x,\,y,\,z$}}
\begin{ruledtabular}
\begin{tabular}{ccccccc}
  &Bond distances (\AA)& \\ \hline
 Cu1-O1	& 1.9854(10)&	C1-C2 &	1.5068(19) \\
 Cu1-O1$^1$ &	1.9853(10)& C2-C3	& 1.3978(19) \\
 Cu1-O4$^2$ &	1.9300(11) & C8-C7 &	1.510(3) \\
 Cu1-O4$^3$	& 1.9300(11) & C3-C5$^4$ &	1.3937(19) \\
 O1-C1 & 1.2777(19)	&  C3-C4 &	1.5061(19) \\
 O2-C1 &	1.2414(19) &	C4-O3	& 1.2350(19) \\
 N1-C8 &	1.490(2) & C4-O4	& 1.2775(18)  \\
 C5-C2	& 1.3926(19) & O4-Cu1$^5$ &	1.9300(11) \\
 C5-C3$^4$	& 1.3937(19) & &	\\	 
 &Bond angles (degree)& \\
 O1$^1$-Cu1-O1&	180.00(5)	 &	C5-C2-C1	& 117.69(12) \\
 O4$^2$-Cu1-O1&	88.47(5)	& C5-C2-C3 &	119.49(12) \\
 O4$^3$-Cu1-O1&	91.53(5)	& C3-C2-C1 &	122.73(13) \\
 O4$^3$-Cu1-O1$^1$&	88.47(5) & N1-C8-C7 &	110.27(16) \\
 O4$^2$-Cu1-O1$^1$&	91.53(5)	& C5$^4$-C3-C2	& 119.36(13) \\
 O4$^2$-Cu1-O4$^3$&	180.00(6)	& C5$^4$-C3-C4	& 118.58(12) \\
 C1-O1-Cu1&	106.39(9)	& C2-C3-C4	& 121.99(12) \\
 C2-C5-C3$^4$&	121.14(13)	& O3-C4-C3	& 119.91(13) \\
 O1-C1-C2&	116.37(13)	& O3-C4-O4	& 126.05(14) \\
 O2-C1-O1&	123.93(14)	& O4-C4-C3	& 114.04(13) \\
 O2-C1-C2&	119.56(13)	& C4-O4-Cu1$^5$ &	124.64(10) \\
\end{tabular}
\end{ruledtabular}
\end{table}

\subsection{Microscopic magnetic model}
\label{sec:microscopic}
\begin{figure}
\includegraphics{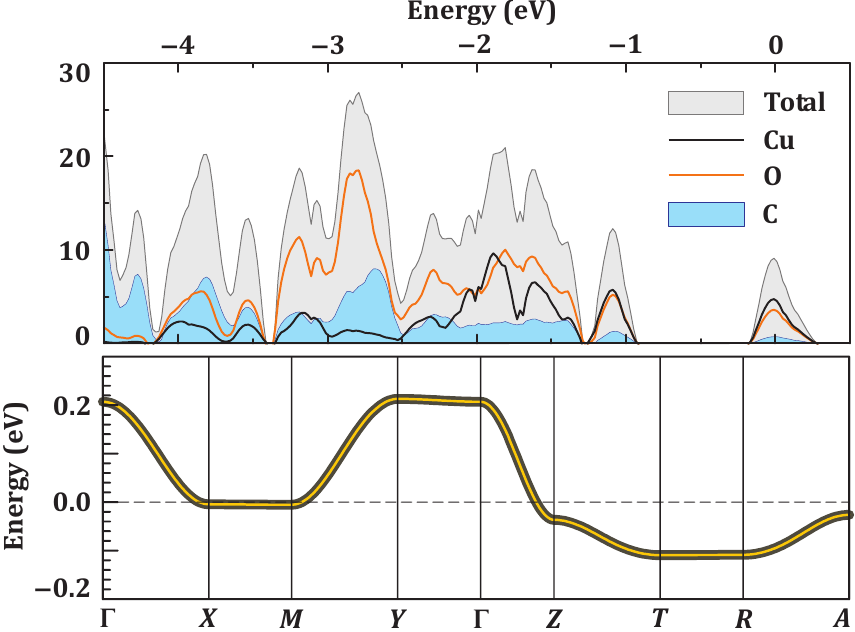}
\caption{\label{fig:dos}
(Color online) Top: GGA density of states (DOS) for Cu(PM)(EA)$_2$. Bottom: the Cu $d_{x^2-y^2}$ band crossing the Fermi level (thin light line) and its tight-binding fit (thick dark line). Note the very weak dispersion along $\Gamma-Y$ and more pronounced dispersions along the two other directions rendering Cu(PM)(EA)$_2$ magnetically quasi-2D.
}
\end{figure}

To determine the magnetic model of Cu(PM)(EA)$_2$, we evaluate individual exchange couplings. This procedure is two-fold. First, we analyze the band structure calculated within GGA. This band structure (Fig.~\ref{fig:dos}, top) is gapless, at odds with the blue crystal color, because essential correlation effects in the Cu $3d$ shell are missing in GGA. Nevertheless, the GGA band structure clearly identifies relevant magnetic states, which are Cu $3d$ orbitals of $x^2-y^2$ symmetry contributing to the single band crossing the Fermi level, as shown in the bottom part of Fig.~\ref{fig:dos} (the $x$ and $y$ axes are directed to the corners of CuO$_4$ plaquettes; they are different from the crystallographic directions $a$ and $b$). The tight-binding description of this band yields hopping integrals $t_i$, which are introduced into a single-band Hubbard model and for the strongly localized case ($t_i\ll U_{\eff}$) at half-filling provide antiferromagnetic (AFM) part of the exchange couplings as $J_i^{\AFM}=4t_i^2/U_{\eff}$, where $U_{\eff}=4.5$\,eV is an effective on-site Coulomb repulsion on the Cu site.\cite{janson2011,tsirlin2013} 

A qualitative inspection of the band in Fig.~\ref{fig:dos} reveals negligible dispersion along $\Gamma\!-\!Y$, hence the magnetic coupling along the $b$ direction is indeed very weak, in agreement with our crystallographic considerations in Sec.~\ref{sec:structure}. On the other hand, comparable dispersions along $\Gamma\!-\!X$ and $\Gamma\!-\!Z$ imply only a weak spatial anisotropy in the $ac$ plane. The evaluation of individual $t_i$'s yields similar values of $t_a$ and $t_c$, whereas second-neighbor couplings $t_2$ and $t_2'$ are very weak. The leading interlayer exchange is $J_{\perp}$ along $[010]$, but it is three orders of magnitude lower than $J_a$ and $J_c$. 

\begin{table}
\caption{\label{tab:exchanges}
Exchange couplings in Cu(PM)(EA)$_2$: metal-metal distances $d_i$ (in\,\r A), electron hoppings $t_i$ of the tight-binding model (in\,meV), AFM contributions to the exchange integrals calculated as $J_i^{\AFM}=4t_i^2/U_{\eff}$ (in\,K), and total exchange integrals $J_i$ from GGA+$U$ (in\,K). Weak exchange couplings $J_2,J_2'$, and $J_{\perp}$ were omitted in GGA+$U$ calculations.
}
\begin{ruledtabular}
\begin{tabular}{ccrcc}
              & $d_i$ & $t_i$ & $J_i^{\AFM}$ & $J_i$ \\ 
  $J_a$       &  5.86 &   37  &  14          &   7   \\
  $J_c$       &  9.18 &   44  &  20          &  10   \\
  $J_2$       & 10.88 &    7  &  0.5         &       \\
  $J_2'$      & 10.90 &   10  &  1.0         &       \\
  $J_{\perp}$ &  8.36 & $-1$  &  0.01        &       \\
\end{tabular}
\end{ruledtabular}
\end{table}

Alternatively, we estimate individual $J$'s from total energies of collinear spin configurations calculated within GGA+$U$. This approach verifies the results of our tight-binding analysis and provides ferromagnetic (FM) contributions to the superexhcnage, which were so far missing from the analysis. The GGA+$U$ results confirm that $J_c>J_a$, and both couplings are slightly below 10\,K. Remarkably, $J_c$ systematically exceeds $J_a$, even though it runs between those Cu$^{2+}$ ions that are further apart. 

Our microscopic analysis concludes that Cu(PM)(EA)$_2$ features a rectangular lattice of Cu$^{2+}$ ions in the $ac$ plane. Both interlayer coupling and frustrating second-neighbor in-plane couplings are very weak. In the following, this microscopic scenario is confirmed experimentally.

\subsection{Magnetization}
\begin{figure}
\includegraphics[scale=0.85]{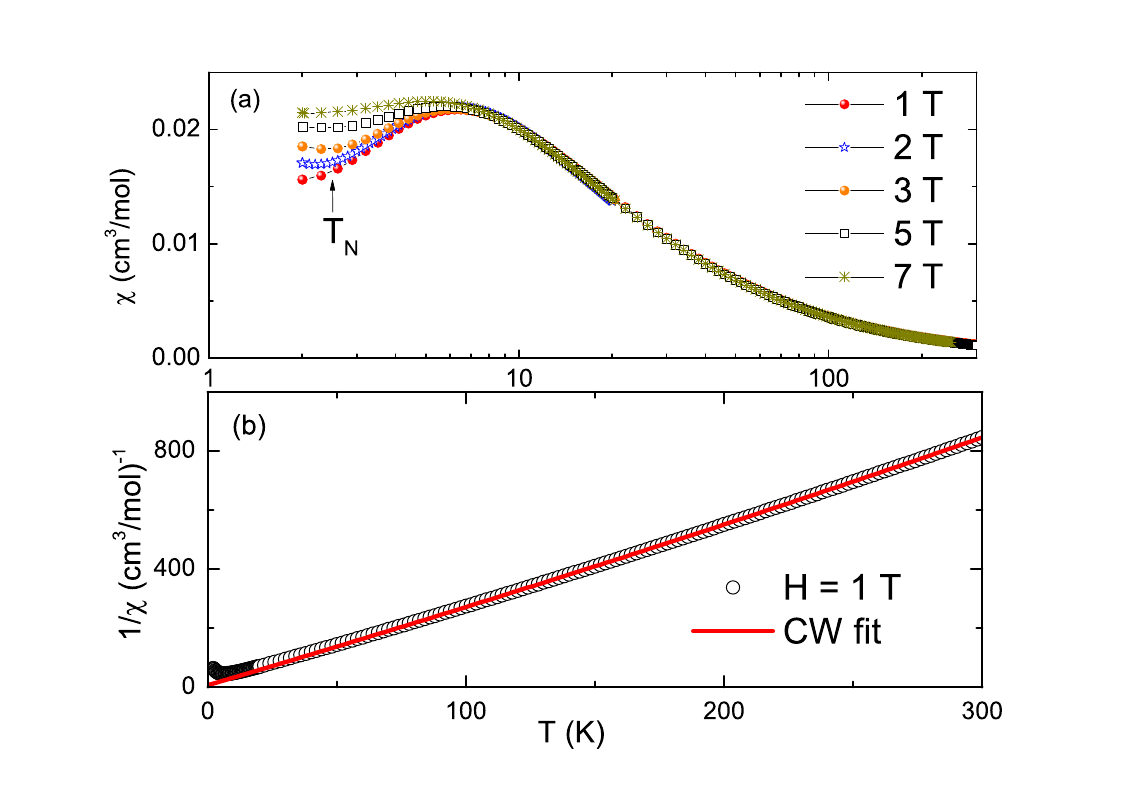}
\caption{\label{chi} 
(Color online) Top: Magnetic susceptibility of Cu(PM)(EA)$_2$ measured in the applied fields between 1 and 7\,T. Bottom: Curie-Weiss fit of the susceptibility data.}
\end{figure}
Magnetic susceptibility $\chi$ as a function of $T$ measured at different applied fields is shown in Fig.~\ref{chi}(a). With decreasing $T$, $\chi(T)$ at 1\,T increases in a Curie-Weiss manner and then shows a broad maximum ($T^{\rm max}_\chi$) at about 6\,K indicative of the short-range magnetic order, which is a hallmark of low-dimensionality. At low temperatures, the susceptibility changes in a smooth manner without any signatures of a magnetic transition, in agreement with the heat-capacity and ESR data reported below.

To fit the bulk susceptibility data at high temperatures, we use the expression
\begin{equation}
\chi=\chi_{0}+\frac{C}{T+\theta_{CW}},
\end{equation}
\newline
where $\chi_{0}$ is the temperature-independent contribution and consists of diamagnetism of the core electron shells ($\chi_{\rm core}$) and Van-Vleck paramagnetism ($\chi_{\rm VV}$) of the open shells of the Cu$^{2+}$ ions present in the sample. The second term is the Curie-Weiss (CW) law with the Curie-Weiss temperature $\theta_{\rm CW}$ and Curie constant $C=N_A\mu_{\rm eff}^{2}/3k_B$, where $N_A$ is Avogadro's number, $k_{\rm B}$ is the Boltzmann constant, $\mu_B$ is the Bohr magneton, and the effective moment is $\mu_{\rm eff} = g\sqrt{S(S+1)} \mu_{\rm B}$/f.u. and f.u. means formula unit. 

Our fit in the temperature range between 210\,K and 300\,K [Fig.~\ref{chi}(b)] yields $\chi_{0}\simeq -1.0\times10^{-4}$\,cm$^{3}$/mol, $C\simeq 0.389$\,cm$^3$\,K/mol, and $\theta_{\rm CW}\simeq 3$\,K. Positive value of $\theta_{\rm CW}$ suggests
that the dominant interactions are AFM in nature. The $C$ value yields an effective moment of 1.76\,$\mu_{B}$, slightly higher than the spin-only $S=\frac12$ value of 1.73\,$\mu_{B}$ (assuming $g=2$) and, thus, corresponding to the $g$-factor above 2.0, which is typical for Cu$^{2+}$ compounds\cite{janson2011,nath2014} and agrees well with the ESR results reported below.

\begin{figure}
\includegraphics{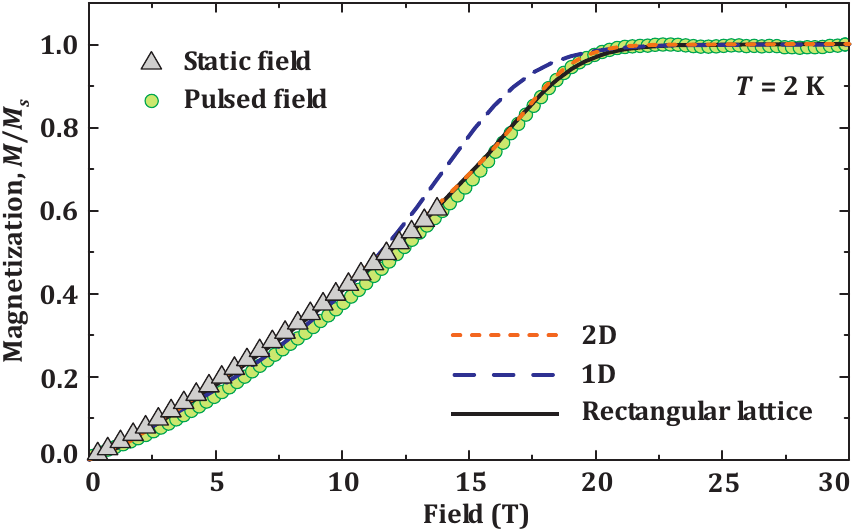}
\caption{\label{MH} Field dependence of the magnetization measured on Cu(PM)(EA)$_2$ in static (triangles) and pulsed (circles) fields. Lines show the simulations with the parameters from Table~\ref{para}, as explained in the text.}
\end{figure}

\begin{table}
\caption{\label{para} 
Parameters obtained from fitting $\chi(T)$ with the rectangular-lattice model ($J_a/J_c=0.7$) as well as purely 1D (uniform chain) and 2D (square lattice) models. $\chi_0$ is the temperature-independent contribution to the susceptibility, $g$ is the $g$-factor, and $J=J_c$ is the exchange coupling. }
\begin{ruledtabular}
\begin{tabular}{crcc}
   & $\chi_0$ (cm$^3$/mol)& $g$ & $J$ (K) \\ \hline
 2D            & $-1.0\times 10^{-4}$ & 2.05 & 6.8  \\ 
 $J_a/J_c=0.7$ & $-1.0\times 10^{-4}$ & 2.07 & 8.0  \\
 1D            & $3.0\times 10^{-5}$  & 2.00 & 10.2 \\
\end{tabular}
\end{ruledtabular}
\end{table}
Magnetization as a function of field is nearly linear in low magnetic fields and reaches saturation at $H_s\simeq 20$\,T (Fig.~\ref{MH}). A slight mismatch between the data measured in static and pulsed fields may be related to dynamic effects. Nevertheless, when scaled against the static-field data, the magnetization in pulsed fields saturates at $M_s\simeq 1.03$\,$\mu_B$/f.u. in excellent agreement with $M_s=gS\mu_B\simeq 1.025$\,$\mu_B$/f.u. expected for $g\simeq 2.05$

\begin{figure}
\includegraphics{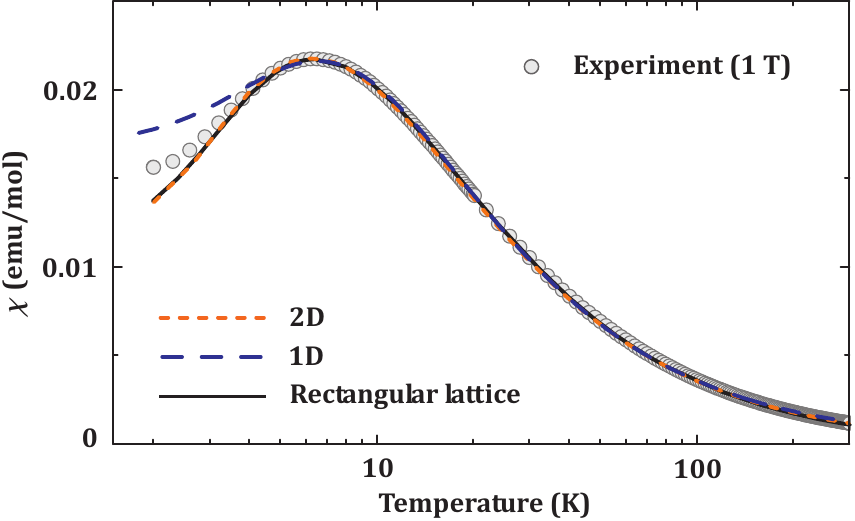}
\caption{\label{model} Fits of the magnetic susceptibility with different spin models: 1D (uniform chain, long-dashed line), 2D (square lattice, short-dashed line), and the rectangular lattice with $J_a/J_c=0.7$ (solid line). Fitting parameters are listed in Table~\ref{para}.
}
\end{figure}
Taking into account the results of the microscopic analysis in Sec.~\ref{sec:microscopic}, we discard frustrated scenarios and focus on the rectangular $J_a-J_c$ spin lattice with $J_a/J_c=0.7$. For the sake of completeness, we also consider the limiting cases of purely 1D ($J_a$ or $J_c$ only) and purely 2D ($J_a=J_c$) spin lattices. First, we fit the susceptibility using $\chi(T)$ obtained from QMC simulations and scaled with the $g$-value, which is a fitting parameter together with the exchange coupling $J=J_c$ and the temperature-independent contribution $\chi_0$. The values of these fitting parameters are listed in Table~\ref{para}. All three models yield fits of comparable quality, although the purely 1D model fails to describe the data in the $3.5-5.0$\,K temperature range, where both 2D models still work reasonably well. As we go from 2D toward 1D, the $J$ value systematically increases because the same overall coupling energy is distributed between only two bonds per site in 1D compared to four bonds per site in 2D. Note also that the $g$-value of $2.05-2.07$ obtained in the fits with the 2D models is in good agreement with ESR (Sec.~\ref{sec:esr}).

The 1D and 2D spin models can be discriminated using high-field magnetization measurements and, in particular, the saturation field $H_s$.\cite{lebernegg2011} In Fig.~\ref{MH}, we show the experimental magnetization curve together with model curves simulated for the parameters from Table~\ref{para}. The purely 1D model yields $H_s^{\rm 1D}=2J\times k_B/(g\mu_B)\simeq 15.2$\,T, which is far below the experimental value. In contrast, the purely 2D model reproduces the experimental curve quite well: $H_s^{\rm 2D}=4J\times k_B/(g\mu_B)\simeq 19.8$\,T. The model of the rectangular lattice is likewise matching the experimental data: $H_s^{\rm rect}=(2J_a+2J_c)\times k_B/(g\mu_B)\simeq 19.6$\,T. Therefore, we conclude that Cu(PM)(EA)$_2$ is clearly a quasi-2D magnet, but the presence of spatial anisotropy in the $ac$ plane (the difference between $J_c$ and $J_a$) can't be assessed from the magnetization data.

\subsection{Heat Capacity}
\label{sec:heat}
\begin{figure}
\includegraphics{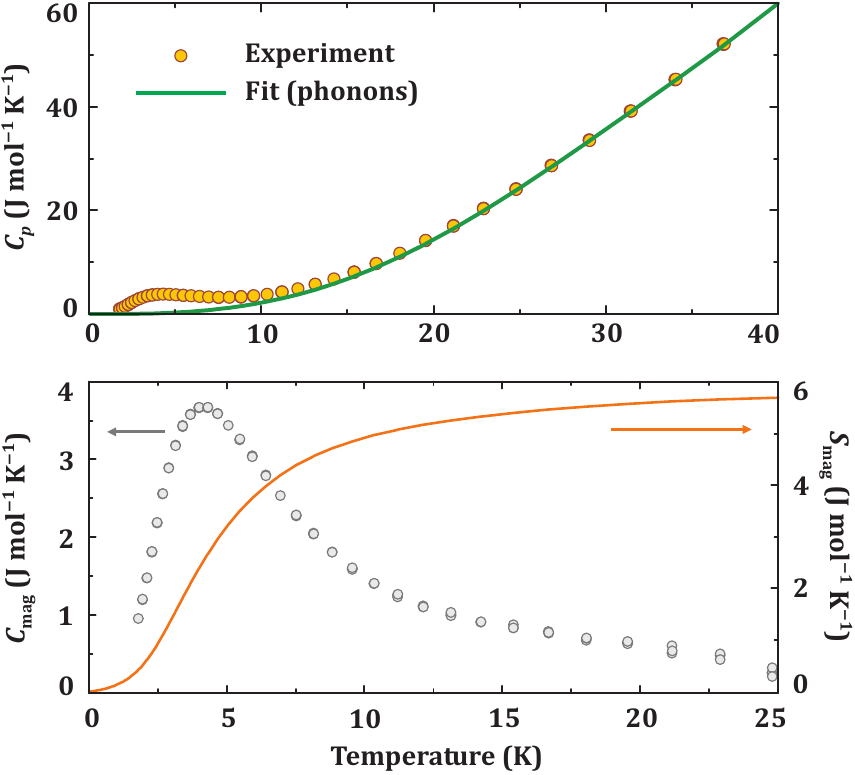}
\caption{\label{cp} Top panel: specific heat ($C_p$) of Cu(PM)(EA)$_2$ measured at zero field as a function of temperature ($T$). Solid line is the fit using Eq.~\eqref{debye} for $T\geq15$\,K with an extrapolation down to 1.8\,K. Bottom panel: magnetic part of the specific heat $C_{\rm mag}$ (solid circles) as a function of $T$. The solid line represents the magnetic entropy $S_{\rm mag}$.}
\end{figure}
A further insight into the nature of Cu(PM)(EA)$_2$ can be obtained from heat-capacity measurements. The heat capacity ($C_p$) in zero field is shown in the top panel of Fig.~\ref{cp}. While at high temperatures it is completely dominated by the contribution of phonon excitations, the magnetic contribution is clearly visible below 10\,K. The maximum around $T_C^{\max}\simeq 4$\,K is characteristic of the short-range order similar to the broad maximum in $\chi(T)$. No kinks associated with the magnetic order are seen down to 1.8\,K.

For a quantitative estimation of $C_{\rm mag}$, the phonon part $C_{\rm phon}$ was subtracted from the total $C_p$. The phonon part was estimated following the procedure used in Refs.~\onlinecite{lancaster2007,matsumoto2000}. Above 15\,K, the data were fitted by the following polynomial
\begin{equation}
C_p(T)=aT^3+bT^5+cT^7+dT^9,
\label{debye}
\end{equation}
where $a$, $b$, $c$, and $d$ are arbitrary constants. \footnote{The fitting results in $a\simeq 0.00474$\,J\,mol$^{-1}$\,K$^{-4}$, $b \simeq -5.95\times 10^{-6}$\,J\,mol$^{-1}$\,K$^{-6}$, $c\simeq 4.0\times 10^{-9}$\,J\,mol$^{-1}$\,K$^{-8}$, and $d\simeq -1.06\times10^{-12}$\,J\,mol$^{-1}$\,K$^{-10}$.} 
The fit was then extrapolated down to 1.8\,K [Fig.~\ref{cp}, top] and subtracted from the experimental $C_p(T)$ data. 

The resulting $C_{\rm mag}(T)$ is shown in the bottom panel of Fig.~\ref{cp}. Its broad maximum is at $T^{\rm max}_C\simeq 4.2$\,K. The subtraction procedure has been verified by calculating the magnetic entropy:
\begin{equation}
S_{\rm mag}(T)=\int_{0}^{T} \frac{C_{\rm mag}(T')}{T'}dT',
\label{entropy}
\end{equation} 
where the data below 1.8\,K were extrapolated with a $T^3$ function.\footnote{The $C_{\rm mag}\sim T^3$ behavior is expected for a 3D antiferromagnet at low temperatures. In the purely 2D case, $C_{\rm mag}$ should be proportional to $T^2$. Both extrapolations yield similar values of $S_{\rm mag}$ with less than 1\,\% difference.}  The estimated $S_{\rm mag}$ at $T=20$\,K is 5.8\,J\,mol$^{-1}$\,K$^{-1}$ in excellent agreement with \mbox{$R\ln[S(S+1)]$} $\simeq 5.76$\,J\,mol$^{-1}$\,K$^{-1}$ expected for $S=\frac12$. Above 20\,K, $C_{\rm mag}$ is very small, and its contribution to the entropy is negligible, hence $S_{\rm mag}$ is nearly constant.

\begin{figure}
\includegraphics{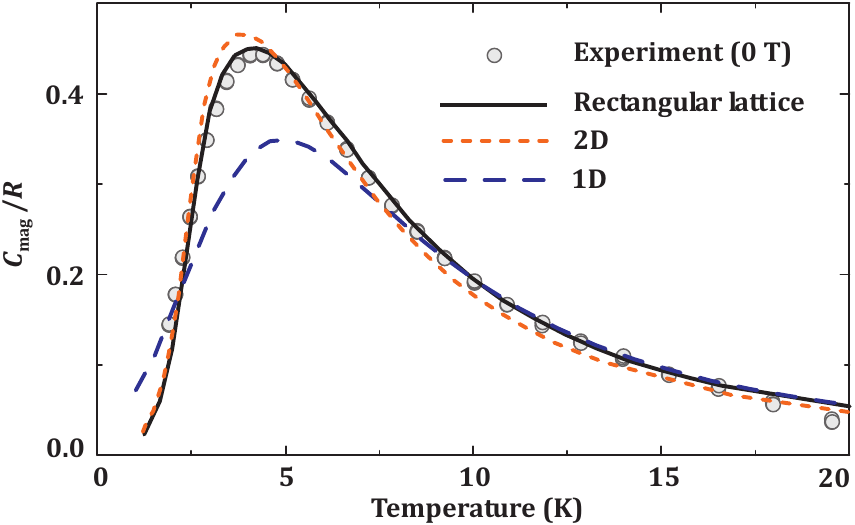}
\caption{\label{cmag} Magnetic contribution to the heat capacity together with QMC simulations for the 1D (uniform chain), 2D (square lattice), and $J_a/J_c=0.7$ rectangular-lattice models.}
\end{figure}
Now, we compare the experimental $C_{\rm mag}(T)$ with simulation results for different spin models (Fig.~\ref{cmag}). Similar to the magnetization data, the purely 1D model utterly fails to reproduce the experiment. The rectangular ($J_a/J_c=0.7$) and square ($J_a/J_c=1$) lattices are again quite similar, although both the exact position and the height of the specific heat maximum clearly favor the rectangular-lattice model. Therefore, we confirm experimentally the weak spatial anisotropy in the $ac$ plane and also demonstrate the remarkable sensitivity of the magnetic specific heat to fine details of the spin lattice.

\begin{figure}
\includegraphics{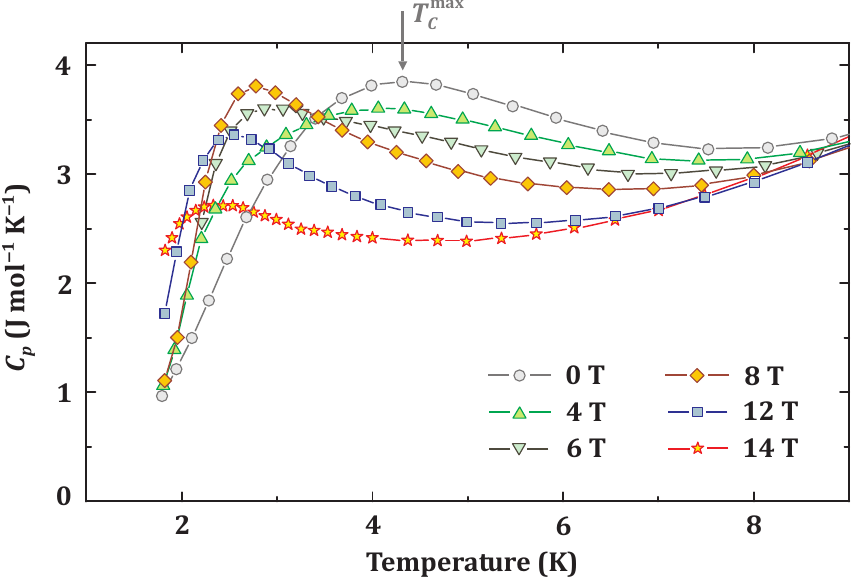}
\caption{\label{TN} Specific heat of Cu(PM)(EA)$_2$ measured in different magnetic fields. The downward arrow points to $T_C^{\rm max}$ in the 0\,T data. In higher fields, $T_C^{\rm max}$ is decreased. }
\end{figure}
Our heat capacity data do not show any signatures of the long-range magnetic order or of any other transition down to 1.8\,K. External field shifts the specific heat maximum toward lower temperatures indicating that a larger amount of magnetic entropy is released at low temperatures when external field is applied. This is typical for quasi-2D antiferromagnets\cite{nath2008,tsirlin2011} because external field suppresses antiferromagnetic spin correlations. The shape of the maximum remains rather symmetric and, thus, distinct from an asymmetric $\lambda$-type anomaly expected at a magnetic transition.

\subsection{Electron spin resonance}
\label{sec:esr}
\begin{figure}
\includegraphics[width=9cm]{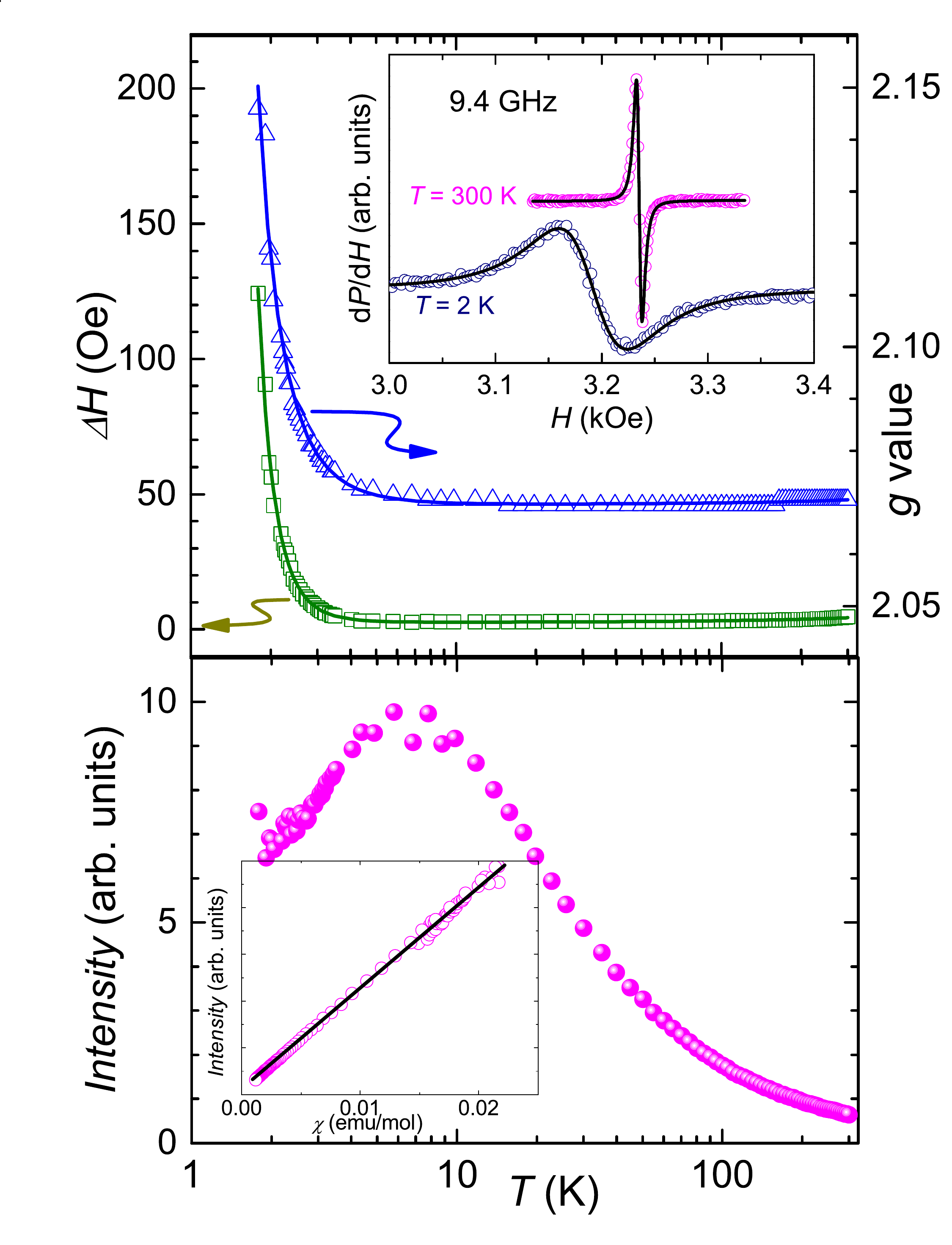}
\caption{\label{ESR}
(Color online) Top: ESR linewidth and $g$ value vs. temperature in the left and right y-axes, respectively. Solid lines are the fits described in the text. Inset: Spectra at $T=300$~K and 2~K and the solid lines are the fits by a Lorentzian function. Bottom: ESR intensity as a function of temperature. Inset: ESR intensity vs. $\chi$ measured at $H=1$~T with temperature as an implicit parameter and the solid line is a linear fit. }
\end{figure}
In order to get additional insight into the low-temperature behavior of Cu(PM)(EA)$_2$, we performed ESR measurements.
ESR measurements on polycrystalline samples revealed a distribution of $g$ values in the range $2.05 \leq g \leq 2.29$ indicating the influence of the square-planar ligand field onto the Cu$^{2+}$ spins resulting in a characteristic $g$ tensor with difference principal values for the in-plane and out-of-plane directions, where the plane is built up by the ligands surrounding the Cu$^{2+}$ ions.\cite{Abragam1970} To investigate the temperature dependence in detail, we were able to orient a small twinned crystal in such a way that the signals of the two main domains merged into a single line. The corresponding direction of the external magnetic field was found to be close to the in-plane case.

The results obtained from this experiment are presented in Fig.~\ref{ESR}.
In the whole measured temperature range, the ESR spectra (inset of the upper panel of Fig.~\ref{ESR}) consist of a single exchange-narrowed resonance line, which is well described in terms of a Lorentz profile. The line width $\Delta H$ is found to be increasing with decreasing temperatures.
As one can see in the upper panel of Fig.~\ref{ESR}, the $g$ value remains close to 2.07 above 10\,K and then starts to diverge below 10~K. Similarly, the ESR line width $\Delta H$ also remains almost constant above 10\,K and then diverges below 10\,K. This low-temperature divergence behavior of $g$ and $\Delta H$ suggests that the compound is approaching magnetically long-range ordered state. In order to extract the parameters associated with the critical divergence, the data were fitted by the power law $(T-T_{\rm N})^{-p}$. Thus we obtained ($T_{\rm N}=0.85(16)$\,K, $p=2.57(34)$) and ($T_{\rm N} = 0.82(12)$\,K, $p=3.89(43)$) from the $g$-value and $\Delta H$ analysis, respectively. This means that the magnetic order is probably approached at temperatures below 0.9\,K. The analysis of the critical behaviour requires measurements below 1.8~K and goes beyond the scope of the present study.

The temperature dependence of the ESR intensity is depicted in the lower panel of Fig.~\ref{ESR}. Calibration of the intensity data using CaCu$_3$Ti$_4$O$_{12}$ as reference,\cite{CCTO} revealed that indeed all copper spins contribute to the ESR signal. The ESR intensity shows a pronounced broad maximum at around 6\,K similar to the $\chi$(T) data. In order to check how the ESR intensity scales with $\chi$, we have plotted intensity vs. $\chi$ with temperature as an implicit parameter in the inset of the lower panel of Fig.~\ref{ESR}. The straight line behavior in the whole temperature range suggests that the ESR intensity tracks the static susceptibility very well, and ESR probes the bulk behavior of the material.

\section{Discussion and Summary}
By combining experimental data with the microscopic analysis, we have shown that Cu(PM)(EA)$_2$ is a non-frustrated quasi-2D antiferromagnet with the weak spatial anisotropy in the $ac$ plane. From thermodynamic properties only, we can't decide which of the couplings in the $ac$ plane is stronger. However, the DFT results convincingly show that $J_c>J_a$, even though the \mbox{Cu--Cu} distance for $J_c$ is nearly twice longer than that for $J_a$ (Table~\ref{tab:exchanges}). This points to the non-trivial nature of the superexchange through the PM anions. To understand the origin of this superexchange process, we explore the nature of ligand orbitals that mix with the half-filled $d_{x^2-y^2}$ orbital of Cu$^{2+}$ and, thus, mediate the superexchange.

\begin{figure}
\includegraphics{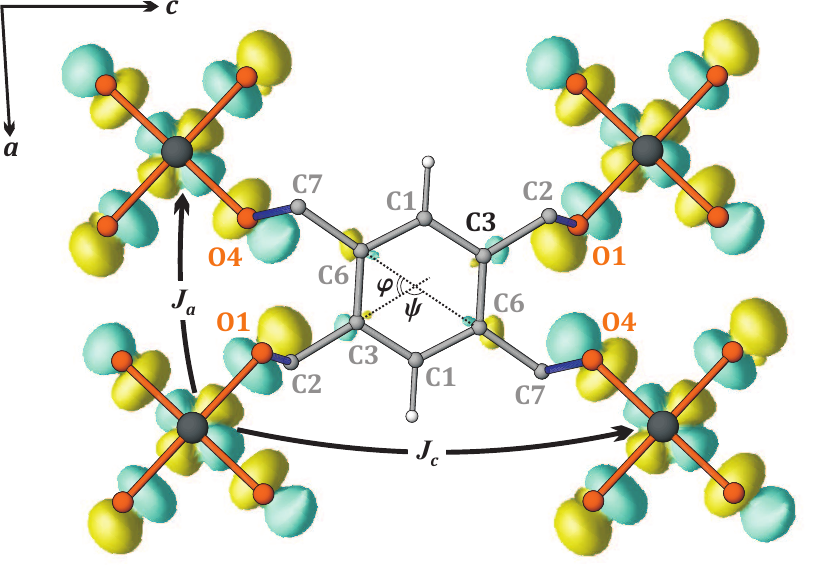}
\caption{\label{fig:wannier}
(Color online) Cu $d_{x^2-y^2}$-based Wannier functions for Cu(PM)(EA)$_2$. Note the ``tails'' of the Wannier functions on the C3 and C6 atoms. Their overlap gives rise to the $J_a$ and $J_c$ superexchange with the effective bridging angles $\varphi\simeq 59.9^{\circ}$ and $\psi\simeq 120.1^{\circ}$, respectively, hence $J_c>J_a$ despite the much longer Cu--Cu distance.}
\end{figure}
The Cu $d_{x^2-y^2}$-based Wannier function (Fig.~\ref{fig:wannier}) features four leading contributions from the $2p$ orbitals of oxygen atoms surrounding the Cu$^{2+}$ ion (O1 and O4). These contributions are about 14\,\% each. Additionally, we find minuscule 2.5\,\% ``tails'' of the Wannier function on the C3 and C6 atoms belonging to the C$_6$ phenyl ring. The difference between $J_a$ and $J_c$ can be now traced back to the positions of relevant $2p$ orbitals on the carbon atoms. Their orientation is fixed by the C2--C3 (C7--C6) bonds, so that the effective bridging angles of the superexchange are $\varphi\simeq 59.9^{\circ}$ and $\psi\simeq 120.1^{\circ}$ for $J_a$ and $J_c$, respectively, and the $J_c$ superexchange is more favorable than that of $J_a$ according to Goodenough-Kanamori-Anderson rules. This explains why the order of magnetic couplings in Cu(PM)(EA)$_2$ does not follow the order of Cu--Cu distances and a counter-intuitive microscopic scenario emerges.

Cu(PM)(EA)$_2$ exhibits an interesting example of the superexchange through a very long \mbox{Cu--O$\ldots$C$\ldots$C$\ldots$O--Cu} pathway. This case is by far more involved than that of Cu(pz)$_2$X$_2$ magnets, where two Cu atoms are directly linked through the pyrazine molecule C$_4$H$_4$N$_2$, with its nitrogen atoms being first neighbors of Cu$^{2+}$ and, thus, featuring large $2p$ contributions to the magnetic orbital.\cite{vela2013} Then the tentative superexchange pathway is Cu--N$\ldots$N--Cu akin to the \mbox{Cu--O$\ldots$O--Cu} pathways that are abundant in Cu$^{2+}$ phosphates and related compounds.\cite{janson2011,nath2014} The interactions of this type are quite sensitive to individual interatomic distances\cite{vela2013} and require that the distance between the ligand atoms (N$\ldots$N or O$\ldots$O) stays below $\sim3.0$\,\r A as to allow for the efficient overlap between the ligand $2p$ orbitals.\cite{tsirlin2013} The case of Cu(PM)(EA)$_2$ is qualitatively different. The stronger coupling $J_c$ pertains to the longer C$\ldots$C distance, hence the spatial arrangement of interacting $2p$ orbitals plays crucial role in this material.

Cu(PM)(EA)$_2$ is a quasi-2D antiferromagnet. It features a non-negligible interlayer coupling $J_{\perp}$ that should trigger long-range magnetic order at low temperatures. Taking our tentative estimate of the interlayer coupling from Table~\ref{tab:exchanges}, we arrive at $J_{\perp}/J_c\simeq 10^{-3}$ and thus expect $T_N/J_c\simeq 0.24$ (Ref.~\onlinecite{yasuda2005}) or $T_N\simeq 2.0$\,K. This temperature is on the verge of our experimental temperature range. Thermodynamic measurements show no evidence for the magnetic order down to 1.8\,K. ESR data suggest that at low temperatures Cu(PM)(EA)$_2$ is approaching the long-range-ordered state with the tentative N\'eel temperature of about 0.85\,K obtained from an empirical fit. While an accurate estimate of the N\'eel temperature requires explicit measurements below 1.8\,K and lies beyond the scope of the present study, even the fact that Cu(PM)(EA)$_2$ does not order down to 1.8\,K is already remarkable and makes this system comparable with the best available quasi-2D antiferromagnets, such as Cu(COO)$_2\cdot 4$H$_2$O and Cu(pz)$_2$X$_2$, where $T_N/J$ is about 0.25.\cite{ronnow2001,lancaster2007} 

Turning now to the in-plane physics, we note that its trends are somewhat counter-intuitive. Within the family of Cu$^{2+}$ square-lattice antiferromagnets, the signatures of magnetic frustration by second-neighbor couplings $J_2$ have been so far observed in Cu(pz)$_2$(ClO$_4)_2$ only.\cite{tsyrulin2009,*tsyrulin2010} In this compound, two nearest-neighbor couplings are mediated by two different pyrazine molecules, hence an efficient superexchange pathway for $J_2$ is missing, because each pyrazine molecule connects nearest-neighbor Cu$^{2+}$ ions only, and any obvious linkage between the second-neighbor Cu$^{2+}$ ions is missing. Our Cu(PM)(EA)$_2$ compound was supposed to remedy this problem by pinning both nearest-neighbor and second-neighbor couplings on the same PM anion. However, it turns out that the superexchange is not mediated by the benzene ring as a whole but by the $2p$ orbitals of individual carbon atoms. The couplings $J_a$ and $J_c$ rely on the orbital overlap between those carbon atoms that are, respectively, first and second neighbors within the hexagonal benzene ring (Fig.~\ref{fig:wannier}). Diagonal couplings $J_2$ and $J_2'$ will, in contrast, require the overlap between third neighbors, which is by far less efficient. 

We speculate that the frustrating coupling $J_2$ can be enhanced by fine-tuning the organic anion. The straight-forward approach of removing two ``idle'' carbon atoms C1 seems to be not viable from chemistry viewpoint. However, five-member rings with a heteroatom, such as the furantetracarboxylic acid C$_4$O(COOH)$_4$, may be suitable molecular bridges for frustrated-square-lattice magnets with comparable first- and second-neighbor couplings. The realm of organic chemistry offers many other acids with cyclic carbon units and four carboxyl-groups (COOH) that are amenable to bond formation with the Cu$^{2+}$ ion. Our work is a natural first step toward the preparation of such quantum magnets and understanding superexchange in these compounds.

In summary, we reported synthesis, crystal structure, magnetic properties, and microscopic magnetic model of a spin-$\frac12$ magnet Cu(PM)(EA)$_2$. Its quasi-two-dimensional magnetic unit features two leading exchange couplings, $J_c\simeq 10$\,K and $J_a\simeq 7$\,K forming a non-frustrated rectangular spin lattice. Superexchange couplings are mediated by carbon atoms of the phenyl ring and conform to the conventional Goodenough-Kanamori-Anderson rules, so that the stronger coupling is $J_c$, even though the relevant Cu--Cu distance is nearly twice larger than that of $J_a$. 

\begin{acknowledgments}
We thank B. R. Sekhar for extending his SQUID-VSM facility at the Institute of Physics, Bhubaneswar for the magnetic susceptibility measurements, as well as Tobias F\"orster and Helge Rosner for their kind help with the high-field magnetization measurements at HLD. RN would like to acknowledge Department of Science and Technology, India for financial support. AAT was supported by the EU under Mobilitas grant MTT77, and by the PUT733 grant of the Estonian Research Agency.
\end{acknowledgments}

\end{document}